\documentstyle[12pt]{article}
\def\p{\hat{p}}
\def\q{\hat{q}}

\def\lb{\label}
\def\be{\begin{equation}}
\def\ee{\end{equation}}
\begin{document}

\title{Multi-loop Feynman Integrals and Conformal Quantum Mechanics
}

\author{
A.P. Isaev
\thanks{e-mail address: isaevap@thsun1.jinr.ru}\\
{\it Bogoliubov Laboratory of Theoretical Physics, JINR, Dubna,} \\
{\it 141 980, Moscow Region, Russia}
}

\date{}

\maketitle

 \begin{abstract}
New algebraic approach to analytical calculations of
$D$-dimensional integrals for
multi-loop Feynman diagrams is proposed.
We show that the known analytical methods of evaluation of
multi-loop Feynman integrals,
such as integration by parts and star-triangle
relation methods, can be drastically simplified by
using this algebraic approach.
To demonstrate the
advantages of the algebraic method of analytical evaluation of
multi-loop Feynman diagrams, we calculate ladder diagrams for the massless $\phi^3$
theory.
Using our algebraic approach we show that the problem of evaluation
of special classes of
Feynman diagrams reduces to the calculation of the Green functions
for specific quantum mechanical problems.
In particular, the integrals for ladder massless diagrams in the $\phi^3$
scalar field theory are given by the Green function for the
conformal quantum mechanics.
 \end{abstract}

\noindent {\bf To the memory of Sergei Gorishnii 1958 - 1988}

\section{Introduction}

It is well known that the evaluation of
 the multiple integrals associated with the Feynman diagrams
is the main source of physical data in the perturbative quantum field theory.
Since the number of diagrams grows enormously in higher orders of perturbation theory,
the numerical calculations of the integrals for multi-loop
Feynman diagrams are not sufficient to obtain results
with desirable precision. That is why analytical calculations of the Feynman diagrams
(integrals) start to be important.

For last few years considerable progress was achieved in analytical
calculations of multi-loop Feynman integrals
(see e.g. \cite{Kreimer} --
\cite{DaKa} and references therein).
It is interesting that in many cases analytical results for the Feynman integrals
are expressed in terms of the multiple zeta values and polylogarithms.
Note that the multiple zeta values and polylogarithms
are very interesting and promising subjects for investigations in
modern mathematics (see e.g. \cite{Zag}, \cite{Gonch}).

The analytical evaluations of the
multi-loop Feynman integrals are usually based on
such powerful methods as the integration by parts \cite{ChTk1} and star-triangle
(uniqueness) relation (see \cite{Kaz}, \cite{Vas} and
references therein) methods. These methods have a long history.
For example, the star-triangle
 relations have firstly been considered in the framework of
the conformal field theories \cite{FP}.
Then it was noticed
\cite{Zam} that the
star-triangle relation is a kind of
the Yang-Baxter equation (see also \cite{Is}).
In \cite{Zam}, this fact was used to calculate the
"fishing-net" Feynman diagram (of a sufficiently large order)
for the four-dimensional ($D=4$) $\phi^4$ theory
(as well as  "triangle-net" and  "honey-comb"
diagrams for the $\phi^6$ ($D=3$) and $\phi^3$ ($D=6$) theories, respectively).

In this paper a new algebraic approach to analytical calculations
of the massless and dimensionally regularized
Feynman integrals, e.g., needed for the renormalization group calculations,
is developed. In particular,
this method is based  on using the integration by parts
and star-triangle (uniqueness) relation methods.
The advantage of our approach is that we change the manipulations with integrals by
the manipulations with the algebraic expressions. This
drastically simplifies all calculations,
as it will be demonstrated by some examples.
In particular, we calculate the integrals for ladder Feynman
diagrams arising in the $\phi^3$ field theory for scalar massless particles.
The integrals for these diagrams contribute to many important physical
quantities and have been extensively used in many applications,
e.g. in the calculations of the conformal
four-point correlators in the $N=4$ supersymmetric Yang-Mills theory
\cite{Sok}, \cite{Ar}. The remarkable fact which we have observed is that
the evaluation of the integrals for special classes of
Feynman diagrams reduces to the calculation of the Green functions
for specific (integrable) quantum mechanical problems.
For example, the integrals for ladder massless diagrams in the $\phi^3$
scalar field theory are given by the expansion over the coupling constant
of the Green function for the D-dimensional
conformal quantum mechanics.

The paper is organized as follows. In Section 2, we
outline the basic concepts of our operator approach. In Section 3,
we explain how the integrals for Feynman diagrams can be
represented in the operator form. The analytical calculations
of the integrals for multi-loop ladder diagrams are presented in
Section 4. In Section 4 we also discuss the relation of multi-loop Feynman
integrals to Green functions of specific quantum mechanical problems.
In Conclusion we discuss possible generalizations and prospects.

\section{Operator Formalism}

Consider the $D$-dimensional Euclidean space $R^D$
with the coordinates $x_i$, where $i=1, \dots, D$.
We denote $(x \, p) = \sum_i x_i \, p_i$,
$x^{2\alpha} = (\sum_i \, x_i \, x_i)^{\alpha} $
and in general the parameter $\alpha$  is a complex number.
Let $\q_i = \q_i^\dagger$ and $\p_i = \p_i^\dagger$ be operators of coordinate and momentum,
respectively,
\be
\lb{gr0}
[\q_k , \, \p_j ] = i \, \delta_{kj}  \; .
\ee
Introduce the coherent states $|x \rangle \equiv |\{ x_i \}\rangle$ and $|k\rangle\equiv |\{ k_i \}\rangle$
which diagonalize the operators of the coordinate and momentum
\be
\lb{gr00}
\q_i |x\rangle  = x_i \, |x\rangle \; , \;\;\;
\p_i |k\rangle  = k_i \, |k\rangle \; ,
\ee
and normalize these states as follows:
\be
\lb{gr1}
\langle x|k \rangle = \frac{1}{(2 \pi)^{D/2}} \, \exp (i \, k_j \, x_j ) \; , \;\;\;
 \int d^D k \, |k\rangle \, \langle k| = \hat{1} = \int d^D x \, |x\rangle \, \langle x| \; .
\ee

We also define the inversion operator $R$
\be
\lb{inver}
\begin{array}{c}
R^2 = 1 \; , \;\;\;
R^\dagger = R \, \q^{2 D} \; , \;\;\;
\langle x | R = \langle {1 \over x} | \; , \\ \\
R \q_i R = \q_i \, / \, \q^2 \; , \;\;\;
K_i = R \p_i R = \q^2 \, \p_i - 2 \, \q_i \, (\q \, \p) \;  .
\end{array}
\ee
where ${1 \over x} := \{ x_i / x^2 \}$.

The well-known formula for the Fourier transformation of
the function $1/k^{2 \beta}$ can be rewritten with the help of eqs.
(\ref{gr00}) and (\ref{gr1})
in the following form:
\be
\lb{gr4}
\langle x| \frac{1}{\p^{2 \beta}} |y\rangle = a(\beta) \,
\frac{1}{(x-y)^{2 \beta'}} \; , \;\;\left( a(\beta) =
\frac{\Gamma(\beta') }{ \pi^{D/2} \, 2^{2\beta} \, \Gamma(\beta)} \right) .
\ee
where $\beta' = D/2 - \beta$, $\Gamma(\beta)$ is the Euler gamma-function
and $\beta' \neq 0, -1, -2, \dots$.
Note that $a(\beta)$ obeys the functional equation
$a(\beta) \, a(\beta') = (2\pi)^{-D}$. Formula (\ref{gr4})
can be interpreted as a definition of the
infinite dimensional matrix representation for $\p^{-2\beta}$
(the matrix "indices" are the coordinates $x_i$ and $y_i$).
In this representation the operator $\q^{2 \alpha}$ is a diagonal matrix
\be
\lb{gr44}
\langle x| \q^{2 \alpha} |y\rangle = x^{2\alpha} \, \delta^D(x-y) \; .
\ee

We call the function $1/(x-y)^{2\alpha}$ the propagator with the index $\alpha$.
Consider the convolution product of two propagators with the indices $\alpha$ and $\beta$:
\be
\lb{gr4b}
\int  \frac{d^D z \,}{(x-z)^{2 \alpha} \, (z-y)^{2 \beta}} =
\frac{V(\alpha' ,\beta')}{(x-y)^{2(\alpha + \beta - D/2)}}
\; ,
\ee
where
$V(\alpha ,\beta) = a ( \alpha + \beta ) / \, (a(\alpha) \, a(\beta) )$.
This is nothing but the group relation
$\p^{-2 \alpha'} \, \p^{-2 \beta'} = \p^{-2 (\alpha' + \beta')}$
which is written in the matrix form (\ref{gr4}).

Using the definition of the inversion operator $R$ (\ref{inver})
one can deduce the main formula
\be
\lb{inver5}
R \, \p^{2 \beta} \, R  =
\q^{2 (\beta + { D\over 2} )} \p^{2 \beta}  \q^{2 (\beta - {D\over 2} )} \; .
\ee
Then, the group relation
$(R \, \p^{2 (\alpha + \beta)} \, R) =
(R \, \p^{2 \alpha} \, R) \, ( R \, \p^{2 \beta} R )$ is equivalent to
the identity
\be
\lb{uniq}
\p^{2 \alpha} \q^{2 \gamma} \p^{2 \beta} =
\q^{2 \beta} \p^{2 \gamma} \q^{2 \alpha} \; , \;\;\;
(\gamma = \alpha + \beta) \; .
\ee
which can be represented in the form of the commutativity condition
$[H_\alpha, \, H_{-\beta}] = 0$
for the operators $H_\alpha = \p^{2 \alpha} \, \q^{2 \alpha}$.
Thus, $H_\alpha$ (for all $\alpha$)
generate the commutative set of elements in the algebra of functions
of $\q_i$, $\p_i$.
Identity (\ref{uniq}), represented in the matrix form (\ref{gr4}), (\ref{gr44}),
is the famous
star-triangle relation $(\alpha + \beta = \gamma)$:
\be
\lb{gr4a}
\int  \frac{d^D z \,}{(x-z)^{2 \alpha'} \, z^{2 \gamma} \,
(z-y)^{2 \beta'} } =
\frac{V(\alpha ,\beta)}{(x)^{2 \beta} \, (x-y)^{2 \gamma'} \,
(y)^{2 \alpha} }
\; .
\ee

Consider the dilatation operator $H := (i/2)(\p \q + \q \p) = - H^{\dagger}$
which satisfies:
 \be
\lb{cartan}
 H \, \p_i = \p_i \, (H - 1) \; , \;\;\;  H \, \q_i = \q_i \, (H + 1) \; ,
 \ee
and generates the $sl(2)$ algebra together with the elements $\q^2$ and $\p^2$:
 \be
 \lb{gr7}
 \begin{array}{c}
[\q^2 , \, \p^2 ] = 4 \, H \; , \;\;\;
[H , \, \q^2] = 2 \, \q^2  \; , \;\;\;
[H , \, \p^2] = -2 \, \p^2   \; .
 \end{array}
 \ee
It is known that the special conformal transformation
generators $K_i$ (\ref{inver}), the dilatation
operator $H$ (\ref{cartan}), the
elements $\p_i$ and $M_{ij} = \q_i \p_j - \q_j \p_i$ generate the
$D$-dimensional conformal algebra $so(D,2)$.

Below the following relations will be important:
\begin{eqnarray}
\lb{is3}
 [ \q^2 , \, \p^{2 (\alpha+1)} ] =
4 \, (\alpha +1) \, (H + \alpha) \, \p^{2 \alpha} \; , \\
\lb{is4}
 [ \q^{2 (\alpha+1)}, \p^2 ] =
4 \, (\alpha +1) \, (H - \alpha) \, \q^{2 \alpha}  \; , \\
\lb{is1}
H \, \q^{2 \alpha} = \q^{2 \alpha}  \, (H + 2 \alpha)  \; , \;\;\;
H \, \p^{2 \alpha}  = \p^{2 \alpha} \, (H - 2 \alpha)  \; .
\end{eqnarray}
These relations are easily deduced from the Heisenberg algebra (\ref{gr0}).
We also introduce the notion of degree for the operators which are
homogeneous functions $\hat{\psi}(\p_i,\q_j)$ of the
generators of the Heisenberg algebra.
The additive number ${\rm deg}(\hat{\psi})$ is called
the degree of the operator $\hat{\psi}$ if
$[\hat{\psi} , \, H ] = {\rm deg}(\hat{\psi}) \, \hat{\psi}$.
In particular, we find
${\rm deg}(\p_i) = - {\rm deg}(\q_j) = 1$ in view of (\ref{cartan}).

Using relations (\ref{is3})
and (\ref{is1}) one can deduce the algebraic identity
\be
\lb{parts}
\begin{array}{c}
4 (2 \gamma - \alpha - \beta ) \,
\p^{2 \alpha} \q^{2 \gamma} \p^{2 \beta} = \\
= \frac{1}{\alpha +1} [\q^2 , \, \p^{2 (\alpha+1)} ] \, \q^{2 \gamma}
\, \p^{2 \beta}
-\frac{1}{\beta +1} \, \p^{2 \alpha} \q^{2 \gamma}
[\q^2, \, \p^{2 (\beta+1)} ] \; .
\end{array}
\ee
In the matrix form (\ref{gr4}), (\ref{gr44}) this identity looks like
\be
\lb{parts0}
\begin{array}{c}
(D- 2\alpha_2 - \alpha_1 - \alpha_3) \, \langle x|\alpha_1, \alpha_2, \alpha_3 |y \rangle = \\
= \alpha_1 \, [ \langle x|\alpha_1^{+}, \alpha_2^{-}, \alpha_3 |y \rangle -
x^2 \, \langle x|\alpha_1^{+}, \alpha_2, \alpha_3 |y \rangle  ] +
\\
+ \alpha_3  \, [ \langle x|\alpha_1, \alpha_2^{-}, \alpha_3^{+} |y \rangle -
y^2 \, \langle x| \alpha_1, \alpha_2, \alpha_3^{+} |y \rangle) ]
\end{array}
\ee
where $\alpha_i^{\pm} = \alpha_i \pm 1$,
$\alpha = -\alpha_1'$, $\gamma = - \alpha_2$, $\beta = -\alpha'_3$
and we have introduced the concise notation for the "vertex" integral
\be
\lb{vertex}
\langle x|\alpha_1, \alpha_2, \alpha_3 |y \rangle =
\int  \frac{d^D z \,}{(x-z)^{2 \alpha_1} \, z^{2 \alpha_2} \,
(z-y)^{2 \alpha_3} } = \frac{1}{a(\alpha_1') a(\alpha'_3)}
\langle x | \frac{1}{\p^{2\alpha_1'}} \, \frac{1}{\q^{2\alpha_2}} \,
\frac{1}{\p^{2\alpha_3'}} | y \rangle
\; .
\ee
Identity (\ref{parts0}) is called
the integration by parts (or ``triangle'') rule \cite{ChTk1}
and plays a very important role in
almost all analytical calculations of multi-loop Feynman integrals.

The integration by parts rule (\ref{parts}) is a combination of
two more fundamental relations. The first one is
 \be
 \lb{difr}
4 (\alpha - \gamma) \p^{2 \alpha} \, \q^{2 \gamma}  =
\frac{1}{\gamma + 1 } \,
\p^{2 \alpha} \, [\q^{2 (\gamma + 1)} , \, \p^2 ]  -
\frac{1}{\alpha + 1 } \, [ \q^2, \, \p^{2 (\alpha +1)} ]
\, \q^{2 \gamma}   \; ,
 \ee
and the second one is obtained from (\ref{difr}) by
the Hermitian conjugation (for real $\alpha$ and $\gamma$).
These relations can also be deduced from
(\ref{is3}) -- (\ref{is1}).

Consider the sequence of products of the operators $\p^{2\alpha_{2k-1}}$,
$\q^{2\alpha_{2k}}$
\be
\lb{seq}
\psi_{\bar{\alpha}} := \psi (\dots ,\alpha_i, \dots) =
 \cdots  \p^{2\alpha_1} \, \q^{2\alpha_2} \, \p^{2\alpha_3}  \, \q^{2\alpha_4} \cdots \; .
\ee
Then the finite difference equation (\ref{difr}) and its conjugated version
can be written in the form of equations $L_i \, \psi (\alpha_i) = 0$
where the finite difference operators
  $$
 \begin{array}{c}
L_i = 4 (\alpha_{i} - \alpha_{i+1}) + \nabla_i + \nabla_{i+1} \\ \\
 \left( \nabla_i =   e^{\partial_{i}} \, \frac{1}{\alpha_{i} } \, \left[
e^{\partial_{i-1}} - e^{\partial_{i+1}}
\right]  \; , \;\; \partial_i := \frac{\partial}{\partial \alpha_i} \right) \; ,
\end{array}
 $$
generate the algebra with the commutation relations:
$[ L_i , \, L_{i+k} ] = 0$ for $(k \geq 3)$ and
$$
[ L_{i-1} - L_i , \, L_{i+1} - L_{i+2} ] = 0 \; , \;\;\;
[ L_i , \, L_{i+2} ] = e^{(\partial_{i+1} +
\partial_{i+2})} \frac{1}{\alpha_{i+1} \alpha_{i+2}}
\, L_{i+1} \; .
$$
This $W$- type algebra has the central element
$Z = -{1 \over 4} \, \sum_j (-)^j \, L_j$ which is
equal to the
degree of the operator (\ref{seq}).

Another set of equations $\sigma_i \, \psi_{\bar{\alpha}} = \psi_{\bar{\alpha}}$ for
the functions (\ref{seq})
follows from the star-triangle identity (\ref{uniq}).
The action of the operators $\sigma_i$ is
\be
\lb{perm}
\begin{array}{c}
\sigma_i \, \psi (\dots \alpha_i, \alpha_{i+1}, \alpha_{i+2},  \alpha_{i+3}, \dots) = \\
= \psi (\dots \alpha_i -\alpha_{i+1}+ \alpha_{i+2}, \alpha_{i+2}, \alpha_{i+1},
\alpha_{i+1}- \alpha_{i+2} + \alpha_{i+3},  \dots) \; .
\end{array}
\ee
The operators (\ref{perm}) define the group
$S_{even} \otimes S_{odd}$ which is a direct product of two symmetric groups
 generated by the sets of even $\{ \sigma_{2i} \}$ and odd $\{ \sigma_{2i+1} \}$
 elements,  respectively. One can show directly that the whole set of the
relations $L_i \, \psi = 0$ and $\sigma_k \, \psi =\psi$
is closed and combinations of these relations do not
lead to new constraints on $\psi$ (\ref{seq}).

It is well known \cite{Hooft} that the dimensional regularization
procedure requires the rule:
$\int \, d^D x / x^{2\alpha} = 0$ for all $\alpha \neq D/2$.
A more precise statement  \cite{GI} is
\be
\lb{gori1}
\int \, d^D x \, \frac{1}{x^{2(D/2 + i \, \alpha)}} = \pi \, \Omega_D \,
\delta(\alpha) \; , \;\;\; (\alpha \in Re)
\ee
where $\Omega_D = \frac{2 \pi^{D/2}}{\Gamma(D/2)}$ is the area
of the unit hypersphere in $R^D$.

In the framework of the dimensional regularization scheme
we extend the definition of the integral in
the left-hand side of (\ref{gori1})
for arbitrary complex numbers $\alpha = |\alpha| e^{i \arg(\alpha)}$ as
\be
\lb{gori}
\int \, d^D x \, \frac{1}{x^{2(D/2 + \alpha)}} = \pi \, \Omega_D \,
\delta( | \alpha | ) \; ,
\ee
where $\delta(| \alpha | )$ is the radial delta-function. It is clear that:
$f(\alpha) \cdot \delta(| \alpha |) = f(0) \cdot \delta(| \alpha |)$
for analytic in $\alpha =0$ functions $f(\alpha)$.

The important consequence of
the definition (\ref{gori})
is that now one can introduce the notion of the trace for the
operators (\ref{seq})
\be
\lb{trace}
Tr (\psi_{\bar{\alpha}} ) =
\int d^D x \,
\langle x| \psi_{\bar{\alpha}} | x\rangle \; .
\ee
It follows from (\ref{gori}) that these traces are proportional
to the delta-function:
\be
\lb{isa}
Tr ( \psi_{\bar{\alpha}} )
=\pi \, \Omega_D \, c_{\alpha_i} \, \delta( | \beta | )
\; , \;\;\; \left( c_{\alpha_i}  = \int d \widetilde{x}
\langle \widetilde{x} | \psi_{\bar{\alpha}} | \widetilde{x}\rangle  \right)
\ee
where $\beta = {1 \over 2} {\rm deg}(\psi_{\bar{\alpha}})$,
$\widetilde{x}_i = x_i/ r$ is a unit vector in $R^D$,
$d^D \, x = \Omega_D \, r^{D-1} \, dr \, d \widetilde{x}$
and $c_{\alpha_i}$ is a coefficient function which
contains the whole information about the correlator
$\langle x| \, \psi_{\bar{\alpha}} \, | x\rangle
= c_{\alpha_i} \, x^{-2(D/2 + \beta)}$.
The simple algebraic arguments which lead to the equation (\ref{isa})
are the following. Note that
$e^{t (H + \frac{D}{2})} \, | x \rangle \, = \, | e^{-t} x  \rangle$
and for $r=e^{t}$ we obtain
$$
\langle x| \psi_{\bar{\alpha}} | x\rangle =
\langle \widetilde{x} | \, e^{t (H - \frac{D}{2})} \,
\psi_{\bar{\alpha}} \, e^{-t (H + \frac{D}{2})} \, | \widetilde{x}\rangle
= e^{-t(2\beta +D)}
\langle \widetilde{x} | \psi_{\bar{\alpha}} | \widetilde{x}\rangle
$$
which is consistent with (\ref{isa}).

\vspace{0.3cm}
\noindent
{\bf Remark 1}.
In the case ${\rm deg}(\psi_{\bar{\alpha}})  = 0$,
the operators $\psi_{\bar{\alpha}}$ (\ref{seq})
(which are cut off by the conditions $\alpha_i =0$ for $i <0$ and $i > 2k$)
are expressed
as a product of the commutative operators $H_\alpha$
and their inverse:
\be
\lb{vtetra1}
\psi_{\bar{\alpha}}
= H_{\alpha_1} \, H^{-1}_{\alpha_1 - \alpha_2} \,
H_{\alpha_1 - \alpha_2 + \alpha_3}  \cdots
H^{-1}_{\alpha_{2k} - \alpha_{2k-1}} \,
H_{\alpha_{2k}} \; , \;\;\; \left( \deg(H_\alpha) =0 \right) \; .
\ee
In this case the problem of calculation of the trace (\ref{trace})
reduces to the spectral problem for the commutative operators $H_{\alpha}$.

\vspace{0.3cm}
\noindent
{\bf Remark 2}. The star-triangle identity (\ref{uniq}) can be generalized to
\be
\lb{star2}
\begin{array}{l}
\sum_{i_1 \dots i_n} \, \left( \p^{2 \alpha}
h^{(1)}_{i_1 \dots i_k}(\p) \right) \, \q^{2 (\alpha + \beta)} \,
h^{(2)}_{i_1 \dots i_n}(\q) \,
\left( \p^{2 \beta} \, h^{(3)}_{i_{k+1} \dots i_n}(\p) \right) = \\ \\
= \sum_{i_1 \dots i_n} \,
\q^{2 \beta} \, h^{(3)}_{i_{k+1} \dots i_n}(\q) \, \left( \p^{2 (\alpha + \beta)} \,
h^{(2)}_{i_1 \dots i_n}(\p) \right) \,
\q^{2 \alpha} \, h^{(1)}_{i_1 \dots i_k}(\q) \; , \;\;\;
\end{array}
\ee
where $h^{(a)}_{i_1 \dots i_k}(.)$ $(a=1,2,3)$ are any tensors
being $k$th-order homogeneous polynomials in $\q_i$ or $\p_i$
($\deg(h_{i_1 \dots i_k}(\p) ) = - \deg(h_{i_1 \dots i_k}(\q) ) = k$) .

\vspace{0.5cm}
\noindent
{\bf Remark 3.} The following relation holds:
\be
\lb{im1}
 \frac{1}{(\q-y)^{2\beta}} \, \frac{a(\alpha)}{\p^{2\alpha'}} \, |x\rangle =
 \frac{1}{(\q-x)^{2\alpha}} \, \frac{a(\beta)}{\p^{2\beta'}} \, |y\rangle \; .
\ee
Indeed,
the contraction of both sides of (\ref{im1}) with an arbitrary state $\langle z|$ and
relation (\ref{gr4}) give the identity.
One deduces from (\ref{im1}) that the vector
$a(\alpha) \, (\q-x)^{2\alpha} \, \p^{-2\alpha'} \, |x\rangle$ is independent of
the parameters $\alpha$ and $x_i$ and, thus, should be canceled
by the operators $\p_i$.

\vspace{0.3cm}
\noindent
{\bf Remark 4}.
The dual analog $\overline{R}$ of the inversion operator (\ref{inver}) can be defined
as
\be
\lb{dinver}
\begin{array}{c}
\overline{R} \p_i \overline{R} = \p_i \, / \, \p^2 \; , \;\;\;
\overline{R} \q_i \overline{R} =
\p^2 \, \q_i - 2 \, \p_i \, (\p \, \q) \; , \\ \\
 \overline{R} \, \q^{2 \beta} \, \overline{R}  =
\p^{2 (\beta + { D\over 2} )} \, \q^{2 \beta}  \, \p^{2 (\beta - {D\over 2} )}\;  .
\end{array}
\ee
where $\overline{R}^2 = 1$, $\overline{R}^\dagger = \overline{R} \, \p^{2 D}$ and
$\langle k | \overline{R} = \langle {1 \over k} |$.

\vspace{0.3cm}
\noindent
{\bf Remark 5}. Certain algebraic identities considered above
(e.g. eqs. (\ref{is3}) and (\ref{is4})) are related to each other by
the obvious $Z_2$ symmetry $\p^2 \leftrightarrow \q^2$,
$H \leftrightarrow - H$ which is the
well known automorphism of the $sl(2)$ algebra (\ref{gr7}).

\section{The Diagrams}

The Feynman diagrams which will be considered
in this paper are graphs with vertices
connected by lines labeled by numbers (indices). With each vertex we
associate the point in the D-dimensional space $R^D$ while the lines of the graph
(with index $\alpha$) are associated with the propagator

\unitlength=1cm
\begin{picture}(25,1.5)

\put(1,0.5){\line(1,0){1.8}}
\put(1.7,0.7){$\alpha$}

\put(0.5,0.5){$x$}
\put(3.2,0.5){$y$}
\put(5,0.5){$=$}
\put(6,0.5){$1 / (x-y)^{2\alpha}$}
\end{picture}
\noindent
The boldface vertices $\bullet$ denote
that the corresponding points are integrated over $R^D$.
These diagrams are called the Feynman diagrams in the configuration space.

The figures and operator form of integral expressions for the
3-point, 2-point diagrams
and the tetrahedron vacuum diagram are

\unitlength=7mm
\begin{picture}(25,4)

\put(-1,1){\line(1,0){6}}
\put(1.7,1.2){$\alpha_3$}
\put(2,3.3){$0$}

\put(1,1){\line(1,2){1}}
\put(0.9,0.9){$\bullet$}
\put(-0.2,1.2){$\alpha_1$}
\put(0.7,2){$\alpha_2$}

\put(3,1){\line(-1,2){1}}
\put(2.9,0.9){$\bullet$}
\put(3.8,1.2){$\alpha_5$}
\put(2.8,2){$\alpha_4$}

\put(-1.2,0.6){$x$}
\put(4.9,0.6){$y$}
\put(5.7,1){$=$}
\put(6.7,1){$ f(\alpha_i) \cdot
\langle x| \p^{-2\alpha_1'} \q^{-2\alpha_2}
\p^{-2\alpha_3'} \q^{-2\alpha_4} \p^{-2\alpha_5'} |y\rangle$}

\put(8,0){\bf Fig.1}

\end{picture}

\unitlength=7mm
\begin{picture}(25,4)

\put(1,2){\line(2,1){2}}
\put(1,2){\line(2,-1){2}}
\put(1.8,2.8){$\alpha_4$}
\put(1.8,1){$\alpha_2$}
\put(3.2,1.9){$\alpha_3$}
\put(0.5,2){$0$}

\put(3,1){\line(0,1){2}}
\put(2.9,0.9){$\bullet$}
\put(2.9,2.9){$\bullet$}

\put(3,3){\line(2,-1){2}}
\put(4,2.8){$\alpha_5$}
\put(3,1){\line(2,1){2}}
\put(4,1){$\alpha_1$}

\put(5.2,2){$x$}
\put(5.9,2){$=$}
\put(6.7,2){$ f(\alpha_i) \cdot
 \langle x| \p^{-2\alpha_1'} \q^{-2\alpha_2} \p^{-2\alpha_3'}
\q^{-2\alpha_4} \p^{-2\alpha_5'} |x\rangle$}

\put(8,0){\bf Fig.2}

\end{picture}

\unitlength=7mm
\begin{picture}(25,4)

\put(0,2){\line(1,0){1.8}}
\put(2.2,2){\line(1,0){1.8}}
\put(1.0,1.65){$\alpha_6$}
\put(0,2){\line(2,1){2}}
\put(0,2){\line(2,-1){2}}
\put(0.8,2.85){$\alpha_4$}
\put(0.8,1){$\alpha_2$}
\put(2.05,2.4){$\alpha_3$}
\put(-0.5,2){$0$}

\put(2,1){\line(0,1){2}}
\put(1.85,0.9){$\bullet$}
\put(1.85,2.9){$\bullet$}
\put(3.85,1.9){$\bullet$}

\put(2,3){\line(2,-1){2}}
\put(3,2.8){$\alpha_5$}
\put(2,1){\line(2,1){2}}
\put(3,1){$\alpha_1$}

\put(4.2,2){$x$}
\put(4.8,2){$=$}
\put(5.5,2){$ f(\alpha_i) \cdot  \int \,
d^D x \, \langle x| \p^{-2\alpha_1'} \q^{-2\alpha_2}
\p^{-2\alpha_3'} \q^{-2\alpha_4} \p^{-2\alpha_5'}
\q^{-2\alpha_6} |x\rangle$}

\put(8,0.5){\bf Fig.3}

\end{picture}

\noindent
where the factor $f(\alpha_i)$
is equal to $(2\pi)^{3D} \, a(\alpha_1) a(\alpha_3) a(\alpha_5)$.

The two-loop propagator-type diagram on Fig. 2 (which is called in \cite{Broad1}
the master two-loop diagram) is obtained from the diagram in Fig. 1 in the case $x=y$.
The expression for the vacuum "tetrahedron" diagram in Fig.3
is the result of the integration of the expression for the master two-loop diagram
with the additional propagator $1/x^{2\alpha_6}$. This expression
is represented in the form of the trace (\ref{trace})
\be
\lb{vtetra}
f(\alpha_i) \cdot
Tr ( \p^{-2\alpha_1'} \q^{-2\alpha_2}
\p^{-2\alpha_3'} \q^{-2\alpha_4} \p^{-2\alpha_5'}
\q^{-2\alpha_6} ) \; .
\ee
The tetrahedral symmetry \cite{GI} of this diagram becomes evident if we
impose, in addition to the cyclic symmetry of the trace ($s_1^3 = 1$):
\be
\lb{cycl}
\begin{array}{l}
s_1: \; Tr ( \p^{-2\alpha_1'} \q^{-2\alpha_2}
\p^{-2\alpha_3'} \q^{-2\alpha_4} \p^{-2\alpha_5'}
\q^{-2\alpha_6} ) = \\ \\
Tr ( \p^{-2\alpha_3'} \q^{-2\alpha_4} \p^{-2\alpha_5'}
\q^{-2\alpha_6} \p^{-2\alpha_1'} \q^{-2\alpha_2} ) \; ,
\end{array}
\ee
the identity (the reflection of the diagram in
Fig.3 with respect to the vertical line; this reflection is equivalent to the
permutation of the vertices $x$ and $0$; $s_2^2 =1$):
\be
\lb{cycl1}
\begin{array}{l}
s_2 : \; Tr ( \p^{-2\alpha_1'} \q^{-2\alpha_2}
\p^{-2\alpha_3'} \q^{-2\alpha_4} \p^{-2\alpha_5'}
\q^{-2\alpha_6} ) = \\ \\
=  \frac{a(\alpha_2) \, a(\alpha_4)}{ a(\alpha_1) \, a(\alpha_5)} \,
Tr ( \p^{-2\alpha_2'} \q^{-2\alpha_1}
\p^{-2\alpha_3'} \q^{-2\alpha_5} \p^{-2\alpha_4'}
\q^{-2\alpha_6} ) \; ,
\end{array}
\ee
The last identity is deduced from (\ref{im1}). With the help of the
transformations $s_1$ and $s_2$ one can permute every two vertices of the
tetrahedron on Fig.3. Thus, the elements $s_1$, $s_2$ generate the
permutation group $S_4$ (e.g. one can check that the element
$(s_1 s_2)$ generates the 4th-order symmetry $(s_1 s_2)^4=1$).

Following the paper \cite{GI} we add to (\ref{cycl}) and (\ref{cycl1})
the symmetry (cf. with (\ref{perm})) deduced from the star-triangle equation
($s_3^2 =1$):
\be
\lb{cycl2}
\begin{array}{l}
s_3: \; Tr ( \p^{-2\alpha_1'} \q^{-2\alpha_2}
\p^{-2\alpha_3'} \q^{-2\alpha_4} \p^{-2\alpha_5'}
\q^{-2\alpha_6} ) = \\ \\
= Tr ( \p^{-2(\alpha_1' -\alpha_2 + \alpha_3' )} \q^{-2\alpha_3'}
\p^{-2\alpha_2} \q^{-2(\alpha_2 -\alpha_3' + \alpha_4)} \p^{-2\alpha_5'}
\q^{-2\alpha_6} ) \; ,
\end{array}
\ee
The symmetries $s_1,s_2,s_3$ (\ref{cycl}) -- (\ref{cycl2})
can be realized as linear transformations of the indices $\alpha_i$ and,
as it was shown in \cite{Broad2}, they
generate the 1440-dimensional finite group $S_6 \otimes Z_2$,
where $S_6$ is a group of permutations of 6 objects and $Z_2$
is a 2-fold cyclic group. The most general analytical result for
the master two-loop diagram on Fig. 2 has been achieved in \cite{BGK} and
\cite{Kot}. According to this result, the trace (\ref{vtetra})
for $\alpha_4 = \alpha_2=1$ can be expressed
in terms of $_3F_2$ hypergeometric series.

At the end of this section we stress that
to write down more complicated diagrams in the operator form,
we need to extend the Heisenberg algebra (\ref{gr0}) to the multi-particle
case
$[\q^{(a)}_{k} , \, \p^{(b)}_{j} ] = i \, \delta_{kj} \, \delta^{ab}$,
where $a,b = 1,2, \dots$ are the numbers of particles.

\section{Applications.}

In this Section we demonstrate how our algebraic methods work
using the example of the $L$-loop ladder ($L$-boxes) diagrams.
 We start with dimensionally
and analytically regularized massless integrals
\be
\lb{br1}
D_L (p_0, p_{L+1}, p; \alpha, \beta, \gamma) =
\left[ \prod_{k=1}^{L} \int
\frac{d^D p_k}{p^{2\alpha_k}_k \, (p_k - p)^{2\beta_k}} \right]
\prod_{m=0}^{L} \frac{1}{(p_{m+1} - p_m)^{2\gamma_m}}
\ee
which correspond to the diagram ($x = p_0$, $y = p_{L+1}$, $z =p$):

\unitlength=6mm
\begin{picture}(25,6)

\put(1,3){\line(1,0){12}}
\put(6.8,5.5){$0$}

\put(2.9,2.9){$\bullet$}
\put(5,2.9){$\bullet$}
\put(3.9,3.2){$\gamma_1$}
\put(1.8,3.2){$\gamma_0$}
\put(4,4.2){$\alpha_1$}
\put(4.7,2.1){$\beta_1$}

\put(5.8,3.2){$\gamma_2$}
\put(6.3,4.1){$\alpha_2$}
\put(6.1,2.1){$\beta_2$}
\put(7,4.1){$\cdots$}
\put(6.5,3.2){$\cdots$}
\put(7,2.1){$\cdots$}

\put(5.2,3){\line(3,4){1.8}}
\put(3,3){\line(5,3){4}}
\put(8.8,3){\line(-3,4){1.8}}
\put(11,3){\line(-5,3){4}}

\put(5.2,3){\line(3,-4){1.8}}
\put(3,3){\line(5,-3){4}}
\put(8.8,3){\line(-3,-4){1.8}}
\put(11,3){\line(-5,-3){4}}

\put(9.7,1.8){$\beta_L$}
\put(9.3,4.1){$\alpha_L$}
\put(9,3.2){$\gamma_{L-1}$}
\put(11.5,3.2){$\gamma_L$}

\put(8.7,2.9){$\bullet$}
\put(10.8,2.9){$\bullet$}

\put(1.2,2.6){$x$}
\put(12.5,2.6){$y$}
\put(6.9,0.2){$z$}

\put(9,0){\bf Fig. 4}

\end{picture}

\noindent
The dual to this diagram
is the ladder $L$-loop diagram for the massless $\phi^3$ theory
(the loops in Fig. 5 correspond to the
boldface vertices in Fig.4):

\unitlength=10mm
\begin{picture}(25,3.3)

\put(1,2.5){\vector(1,0){5}}
\put(6,1){\vector(-1,0){5}}

\put(2,1){\vector(0,1){1.5}}
\put(3.2,1){\vector(0,1){1.5}}
\put(5,1){\vector(0,1){1.5}}
\put(3.5,1.8){$........$}

\put(0.6,2.8){$p_0 - p$}
\put(1.3,0.6){$p_0$}
\put(2.1,2.8){$p_1 - p$}
\put(2.3,0.6){$p_1$}
\put(5.3,2.8){$p_{L+1} - p$}
\put(5.3,0.6){$p_{L+1}$}

\put(1.4,1.8){$p_{10}$}
\put(2.5,1.8){$p_{21}$}
\put(5.3,1.8){$p_{L+1\,L}$}

\put(8,1.8){$(p_{mk} = p_m - p_k)$}

\put(4,0.1){\bf Fig. 5}

\end{picture}

\noindent
This "momentum space" diagram represents the same integral (\ref{br1}),
but in another graphical form.
Using (\ref{gr1}) and (\ref{gr4}) we obtain for (\ref{br1})
the following representation:
\be
\lb{br1a}
D_L (x, y, z; \alpha, \beta, \gamma) =
\left( \prod_{k=0}^L \, \frac{1}{a(\gamma_k')} \right) \,
\langle x| \frac{1}{\p^{2\gamma_0'}} \left( \prod_{k=1}^L \,
\frac{1}{\q^{2\alpha_k}} \frac{1}{(\q-z)^{2\beta_k}}
\frac{1}{\p^{2\gamma_k'}} \right)  |y\rangle \; .
\ee

Now we simplify the problem by fixing indices of the propagators as
$\alpha_k = \alpha$, $\beta_k = \beta$, $\gamma_k = \gamma$,
 and consider the generating function
$$
D_{g} (x, y, z;\alpha , \beta, \gamma ) := a(\gamma')  \,
\sum_{L=0}^{\infty} g^L \, D_L (x, y, z;\alpha , \beta, \gamma ) =
$$
\be
\lb{br6}
=
\langle x \, |  \, \left(  \p^{2\gamma'} - \frac{\bar{g}}{ \q^{2\alpha} (\q-z)^{2\beta}} \right)^{-1}
| \, y \rangle
\ee
where $ \bar{g} = g / a(\gamma \, ')$ is the renormalized coupling constant.
The right-hand side of (\ref{br6}) is the Green function for the two-center operator
$$
{\cal H} = \p^{2\gamma'} - \frac{\bar{g}}{ \q^{2\alpha} (\q-z)^{2\beta}} \; .
$$
This Green function
has the following symmetries under the linear and inverse transformations
of the coordinates $(x,y,z)$:
\be
\lb{br1aa}
D_g (x, y, z; \; \alpha, \beta,  \gamma) =
D_g (y, x, z; \; \alpha , \beta, \gamma ) \; ,
\ee
\be
\lb{br1ab}
D_g (x - z, y - z, -z; \; \beta, \alpha, \gamma) =
D_g (x, y, z; \; \alpha , \beta, \gamma ) \; ,
\ee
\be
\lb{br1b}
D_{g z^{-2\beta}} ({1 \over x}, {1 \over y}, {1 \over z}; \; \widetilde{\alpha}, \beta, \gamma) =
x^{2\gamma} \, y^{2 \gamma} \,
D_g (x, y, z; \; \alpha , \beta, \gamma ) \; .
\ee
Here $1 / x := \{ x_i / x^2 \}$ and
\be
\lb{br1bb}
\widetilde{\alpha} = 2 \, \gamma' - \alpha - \beta =
D -  ( \alpha + \beta + 2 \, \gamma ) \; .
\ee
The proof of (\ref{br1aa}) and (\ref{br1ab}) is evident.
The symmetry (\ref{br1b}) follows from the chain of equalities
$$
\langle x \, |  R^2 \,
\left(  \p^{2\gamma'} - \frac{\bar{g}}{ \q^{2\alpha} (\q-z)^{2\beta}} \right)^{-1} | \, y \rangle =
$$
\be
\lb{chain}
=
\langle x \, |  R \,
\left(  \q^{2(\gamma' + D/2)} \,
\p^{2\gamma'} \, \q^{2(\gamma' - D/2)}
- \frac{\bar{g}z^{-2\beta} }{
\q^{-2(\alpha + \beta)} (\q-{1 \over z})^{2\beta}} \right)^{-1} \, R \, | \, y \rangle =
\ee
$$
= x^{-2\gamma} \, y^{-2\gamma} \, \langle {1 \over x} \, |  \,
\left( \p^{2\gamma'}
- \frac{\bar{g}z^{-2\beta} }{
\q^{\, 2\widetilde{\alpha}} \, (\q-{1 \over z})^{2\beta}} \right)^{-1} \, | \,{1 \over y} \rangle
$$
where equations (\ref{inver}) and (\ref{inver5}) where applied.

Using identities (\ref{br1ab}) and (\ref{br1b}) one can change the indices of the
propagators in (\ref{br1}).
In the case $\widetilde{\alpha} = 0$ (see (\ref{br1bb}))
the combination of the symmetries (\ref{br1ab}) and (\ref{br1b}) gives
\be
\lb{br5}
D_g (x, y, z; \; \alpha , \beta , \gamma ) =
 x^{-2\gamma} \, y^{-2\gamma}  \,
G_{\gamma',\beta}(u,v)
\ee
where
\be
\lb{br7}
G_{\gamma \, ',\beta}(u,v)
= \langle u \, |  \, \left( \p^{2\gamma \, '}
- \frac{ g_z }{ \q^{2\beta} } \right)^{-1} \, | \, v \rangle \; ,
\ee
and
$$
g_z = \bar{g}z^{-2\beta}= \frac{g }{ a(\gamma \, ')  \, z^{2\beta} } \; , \;\;\;
u_i =  {x_i \over x^2}- {z_i \over z^2} \; , \;\;\;
v_i = {y_i \over y^2}- {z_i \over z^2} \; .
$$

The function $G_{\gamma \, ',\beta}(x,y)$
is obviously related to the function (\ref{br6})
(if we put there $\beta = 0$, $\bar{g} = g_z$), and for the case $g_z = const$
 from (\ref{br1aa}), (\ref{br1b}) we obtain the symmetry
\be
\lb{br8}
G_{\gamma,\beta}(u,v) = G_{\gamma,\beta}(v,u) = (u^2 v^2)^{(\gamma - {D \over 2})} \,
G_{\gamma,2\gamma -\beta} ({1 \over u}, {1 \over v}) \; ,
\ee
(note that here we have made the redefinition:  $\gamma \, ' \rightarrow \gamma$).

Thus, the problem of analytical evaluation
of the integrals (\ref{br1}) for the ladder diagrams
with the special choice of the line indices
$( \alpha + \beta + 2 \, \gamma = D)$ is reduced
to the problem of calculation of the Green function
for the Hamiltonian
$$
{\cal H} = \p^{2\gamma \, '} - \frac{ g_z }{ \q^{2\beta} } \; .
$$
which describes (for $\gamma' =1$)
the propagation of the scalar particle in the external field of one
source fixed in the origin.
Now the most interesting cases are: \\
1.) $\gamma' = \beta = 1$, $(\alpha =1)$ which corresponds to the $D$-dimensional
conformal mechanics; \\
2.) $\gamma' = 1$, $\beta = D/2 - 1$, $(\alpha =3 - {D \over 2})$
which is related to the standard $D$-dimensional problem
of the dynamics of a
charged particle moving in the field of the pointlike source. For $D=4$ both the
cases are equivalent to each other and the Green function (\ref{br7})
defines the generating function for the ladder diagrams in the $\phi^3$ $(D=4)$ field theory.

\vspace{0.3cm}
Here we calculate the Green function (\ref{br7}) and, therefore,
the integral (\ref{br1}) for general value of $D$
and for the choice of the indices $\gamma = D/2 -1$, $\alpha = \beta = 1$
($D$-dimensional conformal mechanics) with the help of the operator approach.
Our method is based on the identity:
\be
\lb{grn1}
\frac{1}{\p^2 -  g / \q^2 }  =
\sum^\infty_{L=0}
\left( -\frac{g}{4} \right)^L \,  \left[ \q^{2\alpha} \,
\frac{(H-1)}{  (H-1+ \alpha)^{L+1}} \, \frac{1}{\p^2} \,
\q^{-2\alpha}  \right]_{\alpha^L}
\ee
where we denote
$[ \dots ]_{\alpha^L} = \frac{1}{L!} \,
\left( \partial_\alpha^L \,
\left[ \dots \right] \right)_{\alpha = 0}$.
Identity (\ref{grn1}) can be proved immediately if one acts on both the sides
by ${\cal H} = \p^2 -  g / \q^2$ and uses relations (\ref{is3})-(\ref{is1})
and $\left[ (h - 2\alpha)(h -\alpha)^{-L-1} \right]_{\alpha^{L}}
 = \delta_{L,0}$, $\forall h \neq 0$.
Taking into account the integral representation for the rational function of $H$
in (\ref{grn1})
$$
\frac{(H-1)}{  (H-1+ \alpha)^{L+1}} = \frac{(-1)^{L+1}}{L!} \,
\int^\infty_0 \, dt \, t^L \, e^{t\alpha}
\, \partial_t \left( e^{t \, (H-1)} \,  \right)
$$
the Green function $G_{1,1}(u,v)$ (\ref{br7}) is written in
the form
\be
\lb{grn2}
\langle u| \, \frac{1}{(\p^{2} - g_z/\q^{2})} \, |v\rangle =
 \sum^\infty_{L=0}
\frac{1}{L!} \left( {g_z \over 4} \right)^L \, \Phi_L(u,v)
\ee
where for the functions $\Phi_L(u,v)$ we obtain the integral representation
$$
\Phi_L(u,v) =
- \int^\infty_0 \, dt \, t^L \left[ \langle u| \left( \q^{2} \, e^{t} \right)^\alpha
\, \partial_t \left( e^{t \, (H-1)} \right)\, \frac{1}{\p^2}
\q^{-2\alpha} |v\rangle  \right]_{\alpha^L} =
$$
\be
\lb{grn3}
= - a(1) \, \int^\infty_0 \, dt \, t^L \left[
\left( \frac{u^{2} }{v^2} \right)^\alpha \, e^{t\alpha} \right]_{\alpha^L}
\,  \partial_t \left(
\frac{ e^{-t}}{\left(u - e^{-t}v \right)^{2} } \right)^{({D \over 2}-1)}
\ee
Note that the first function is:
$\Phi_0 = a(1) \, (u -v)^{2(1 - {D \over 2})}$ (as it also follows
from (\ref{grn2})) and
in view of (\ref{br8}) all $\Phi_L$ possess the symmetry
$$
\Phi_L(u,v) = \Phi_L(v,u) =  (u^2 v^2)^{(1 - {D \over 2})} \,
\Phi_L ({1 \over u}, {1 \over v}) \; .
$$

The integral (\ref{grn3}) for $D=4$ reproduces the results of
\cite{DU1} and \cite{Bro2}. For applications it is worth having the
general expression (\ref{grn3}) for $D=4-2\epsilon$ and
expand $\Phi_L(u,v)$ over small $\epsilon$:
\be
\lb{expan}
\Phi_L(u,v) = \frac{\Gamma(1-\epsilon)}{4\pi^{2-\epsilon} u^{2(1-\epsilon)} }
\sum_{l=0}^{\infty} \frac{\epsilon^l}{l!} \, \Phi^{(l)}_L(z_1,z_2) \; .
\ee
Here the dimensionless parameters $z_1,z_2$ are defined
by the equations: $z_1 + z_2 = 2(uv)/u^2$ and $z_1z_2 = v^2/u^2$.
The coefficient functions $\Phi^{(l)}_L$ in the expansion
(\ref{expan}) can be represented in the
following form:
\be
\lb{coef}
\Phi^{(l)}_L(z_1,z_2) =
 \sum_{f=0}^{L} \, \frac{(-\ln(z_1z_2))^f \, (2L-f)}{ f! \, (L-f)!}
 \sum_{m=0}^l (-)^m \, C^m_l \,
 {\bf Z}_{m}\left(z_1,z_2;2L +l-f \right) \; ,
\ee
 where $C^m_l = {l \, ! \over m! \, (l-m)!}$ and
 we denote $(k > m)$:
 $$
 {\bf Z}_{m}(z_1,z_2;k) =
\frac{\Gamma(k-m)}{(z_1 -z_2) } \; \sum_{\{n_i\} =1}^{\infty}
\frac{(z_1^{n_0} - z_2^{n_0})}{\left( \sum_{0}^m n_i \right)^{k-m} }
 \left(\prod^m_{i=1} \frac{z_1^{n_i} + z_2^{n_i}}{n_i} \right) \; .
$$
For example, we have
$$
\begin{array}{c}
 {\bf Z}_{0}(z_1,z_2;k) =
 \frac{\Gamma(k)}{z_1-z_2} \, \left( {\rm Li}_k(z_1) - {\rm Li}_k(z_2)\right) \; , \\ \\
  {\bf Z}_{1}(z_1,z_2;k) =
 \frac{\Gamma(k-1)}{z_1-z_2} \, \left( {\rm Li}_{k-1,1}(z_1,1) +
 {\rm Li}_{k-1,1}(z_1, {z_2 \over z_1} ) - (z_1 \leftrightarrow z_2) \right) \; ,
 \end{array}
$$
and in general the functions ${\bf Z}_{m}(z_1,z_2;k)$ are
linear combinations of multiple polylogarithms
$$
{\rm Li}_{k-m,m_1, \dots, m_r}(w_0,w_1, \dots , w_r) =
\sum_{n_0 > n_1 > \dots > n_r > 0} \: \frac{w_0^{n_0} w_1^{n_1} \cdots w_r^{n_r}}{
n_0^{k-m} n_1^{m_1} \dots n_r^{m_r} }
$$
with $\sum m_i = m$ and arguments $w_i$ belong to the set
$\{1,z_1,z_2, {z_1 \over z_2}, {z_2 \over z_1}, {z^2_1 \over z^2_2} \dots \}$.
The first coefficient function in (\ref{coef}) (related to the
integral (\ref{grn3}) for $D=4$)
\be
\lb{d4}
\Phi^{(0)}_L(z_1,z_2) =  \frac{1}{z_1-z_2} \, \sum_{f=0}^{L} \,
\frac{(-)^{f}  \, (2L-f)!}{ f! \, (L-f)!} \, \ln^f(z_1z_2) \,
\left[ {\rm Li}_{2L-f}(z_1) - {\rm Li}_{2L-f}(z_2)\right] \; ,
\ee
has firstly been evaluated in \cite{DU1}.
The next coefficient is
$$
\Phi^{(1)}_L(z_1,z_2) =
 \sum_{n=L}^{2L} \, \frac{n \, (-\ln(z_1z_2))^{2L-n} }{ (2L-n)! \, (n-L)!}
\left( {\bf Z}_{0} (z_1,z_2;n+1) - {\bf Z}_{1} (z_1,z_2;n+1) \right) \; ,
$$

We also present the result for the integral (\ref{grn3})
in the case $u=v$ $(x=y$) and arbitrary $D < 2(L+1)$ which is needed for
evaluation of the propagator-type ladder diagrams:
$$
u^{2({D \over 2}-1)} \, \Phi_L(u,u)
= \frac{a(1) (2L-1)!!}{  \Gamma(D-2)} \, \sum_{k=0}^{\infty}
\frac{\Gamma(k+D-2)}{k! \, (k +D/2 -1)^{2L} }  \; .
$$
For $D=4$ we immediately obtain $u^{2} \, \Phi_L(u,u) \sim  \zeta(2L-1)$.

\vspace{0.5cm}
\noindent
{\bf Remark 1.} Relation (\ref{is4}) can be rewritten in the form
$$
\frac{1}{\p^2} \, \q^{2(\alpha -1)} \, \frac{1}{\p^2} =
\frac{-1}{4\alpha (H-\alpha -1)} \left[ \q^{2\alpha} , \, \frac{1}{\p^2} \right] =
\frac{1}{4\alpha} \, \int_0^{\infty} dt \, e^{t(H-\alpha-1)} \,
\left[ \q^{2\alpha} , \, \frac{1}{\p^2} \right] \; .
$$
In the matrix representation (\ref{gr4}), (\ref{gr44})
this identity gives the analytical
expression for the vertex integral (\ref{vertex}) for the special choice
of the indices (one index is arbitrary)
\be
\lb{vert2}
\langle x|{D \over 2} -1, 1-\alpha, {D \over 2} -1 |y \rangle =
\frac{1}{4 \alpha a(1)} \,  \int_0^{\infty} dt \, e^{t({D \over 2} -1)} \,
\frac{ (x^{2} e^t)^\alpha - (y^{2} e^{-t})^\alpha}{(e^t x-y)^{2({D \over 2} -1)} }
\ee
It reproduces
the results presented in \cite{Usyuk}, \cite{Davyd2} since the indices of the lines
in the left-hand side of (\ref{vert2})
are changed by means of the star-triangle symmetries (\ref{perm}).

\vspace{0.5cm}
\noindent
{\bf Remark 2.}
The one- and two-loop ladder integrals (\ref{br1})
for the case $D=4$ and $\alpha_k = \beta_k = \gamma_k =1$
(which are given by the functions $\Phi^{(0)}_1$ and $\Phi^{(0)}_2$ (\ref{d4})) have been
found in \cite{DU2}. This result was generalized in \cite{DU1} and \cite{Bro2}
where (\ref{d4}) for all $L$
and the generating function (\ref{grn2})
(for the case $D=4$ and $\alpha_k = \beta_k = \gamma_k =1$)
has been calculated. The analytical results for
1,2,3-loop (boxes) integrals (\ref{br1}), general value of $D$ and
special indices of the lines
$\alpha_k,\beta_k,\gamma_k$ can be found in
\cite{Smir1}, \cite{Davyd2},
\cite{Smir2}, \cite{Rem}, \cite{Taus} (see also references therein).

\vspace{0.5cm}
\noindent
{\bf Remark 3.}
In the case
of the conformal indices: $\alpha_k + \beta_k + \gamma_k +\gamma_{k-1} =D$ $(\forall k)$
the inversion method (\ref{chain}) reduces the 4-point functions
(\ref{br1}), (\ref{br1a}) (represented on Fig. 4) to the 3-point functions.
Moreover, the inversion method (\ref{chain}) gives the possibility to obtain the
duality transformation for
the $D$-dimensional Green function
$$
\langle x| \frac{1}{\p^{2\alpha} + g \, (\q^2)^{-2\alpha} +E } | y \rangle
= (x^2 y^2)^{(\alpha - {D \over 2})}
\langle {1 \over x} \, | \, \frac{1}{ \p^{2\alpha} + g  + E \, (\q^2)^{-2\alpha}}
  \, | \, {1 \over  y} \rangle
$$
which relates (for the fixed parameters $\alpha =1,E=0$)
the $D$-dimensional Green function for the Hamiltonian
${\cal H} = \p^2 + g \, (\q^2)^{-2}$ to the propagator of a
massive ($m^2 =g\mu^2$) free particle
(the dimension parameter $\mu$ can be restored by the renormalization
$\p \rightarrow \p/\mu$ and $\q \rightarrow \mu \q$).

The inversion method (\ref{chain}) in the momentum representation
(with the help of the operator $\overline{R}$ (\ref{dinver}))
gives the duality transformation for the Green function
$$
(k^2 \, \tilde{k}^2)^{(\alpha + D) \over 2} \,
\langle k| \left( \p^{2\alpha} + \frac{g}{ (\q^2)^{{\alpha \over 2}}} + E \right)^{-1} | \tilde{k} \rangle =
 \langle {1 \over k} \, |
 \left( 1 + \frac{g}{ (\q^2)^{{\alpha \over 2}}} +
 E \, \p^{2\alpha} \right)^{-1}  | \, {1 \over \tilde{k}}\rangle \; ,
$$
where $g$ and $E$ are dimensionless parameters. For $\alpha =1,D=3$
we have deduced the duality transformation for the
Green function of the Coulomb problem.

\section{Conclusion}

In this paper
the algebraic approach to analytical calculations of the integrals for
multi-loop Feynman diagrams is developed. Possible modifications and
applications of the
approach are the following. First of all we plan to
extend the method to the case of massive propagators.
It is rather trivial to generalize the integration by parts identities
(\ref{parts}), (\ref{difr}) to the massive case. The
massive generalization of the
star-triangle identities (\ref{uniq}), (\ref{star2}) is unknown
(see, however, Sect. 3.2 in the paper \cite{Davyd1}).

The algebraic formulation of the Gegenbauer polynomial technique
\cite{ChTk}, \cite{Kot} is also
possible and based on the identity
\be
\lb{gegen}
\frac{1}{a(\alpha ')} \,  \frac{1}{\p^{2 \alpha'}} =
\sum_{n=0}^{\infty} \, \q^n \left[ \int_{w^2 > z^2}  d^D z \, d^D w \; |z \rangle
 \; C^{\alpha}_n(\tilde{z}\tilde{w}) \; \langle w| + (w \leftrightarrow z)
 \right]  \, \q^{-n-2\alpha} \; ,
\ee
where $\tilde{z}\tilde{w}= (wz)/(w^2z^2)^{1 \over 2}$
and $C^\alpha_n(t)$ are Gegenbauer polynomials.
It is tempting to apply (\ref{gegen}) for the evaluation of
 the two-loop master diagram on Fig. 2.

Finally we stress  that our methods resemble the methods used
in the theory of the quantum integrable systems. For example,
the star-triangle identity (\ref{gr4a}) is a kind of the
Yang-Baxter equation \cite{Zam}, \cite{Is} and can be used for the
$R$-matrix formulation \cite{Derkach} of the integrable
noncompact Heisenberg spin chain
proposed by Lipatov \cite{Lip} for describing the high-energy scattering
of hadrons in QCD. So we hope that
the algebraic approach developed in this paper will be helpful
in investigations of the Lipatov's model.

\vspace{0.5cm}
\noindent {\bf Acknowledgments.}
I would like to thank G. Arutyunov, D. Broadhurst, A.I. Davydychev, H. Gangl,
D.I. Kazakov, A.V. Kotikov, L.N.~Lipatov,
O.V.~Ogievetsky, V.A.~Smirnov, F.V.~Tkachov, A.A.~Vladimirov and A.B.~Zamolodchikov
for interesting discussions and comments. I also
thank the Max-Planck-Institut f\"{u}r Mathematik
in Bonn, where in the beginning of 2002
the considerable part of this work was done, for their kind hospitality
and support.

\end{document}